\documentclass[final]{aipproc}
\layoutstyle{8x11single}

\begin{document}

\title 
[Optical observational biases in the GRB  redshift]
{Optical observational biases in the GRB  redshift}

\classification{ 95.85.Pw,  95.75.Pq, 98.70.Rz}
\keywords {$\gamma$-ray sources; gamma ray burst; statistical analysis}

\author{Z. Bagoly}{address={Dept. of Physics of Complex Systems, E\" otv\" os University, H-1117 Budapest, P\'azm\'any P. s. 1/A, Hungary}}
\iftrue
\author{P. Veres}{address={Dept. of Physics of Complex Systems, E\" otv\" os University, H-1117 Budapest, P\'azm\'any P. s. 1/A, Hungary},altaddress={ Dept. of Physics, Bolyai Military University, H-1581 Budapest, POB 15, Hungary}}

\fi

\copyrightyear  {2008}

\begin{abstract}
The measured redshifts of gamma-ray bursts (GRBs), which were first detected by
the Swift satellite, seem to be bigger on average than the redshifts of GRBs
detected by other satellites.  We analyzed the redshift distribution of GRBs
triggered and observed by different satellites (Swift\citep{sak08}, HETE2\citep{van04}, BeppoSax,
Ulyssses). After considering the possible biases {significant difference was
found at the $p=95.70\% $ level in the redshift distributions of GRBs measured
by HETE and the Swift. }
\end{abstract}

\date{\today}

\maketitle

\section{Introduction}

In this paper we extend our work \cite{2006A&A...453..797B} where we compared
the redshift data of the GRBs triggered by the Swift and by other non-Swift
spacecrafts.  In that paper five statistical tests showed $p \ge 99.40$\%
significance comparing the redshift distributions for the Swift and non-Swift
samples between 01/01/2005-31/01/2006, suggesting that the redshifts of the
Swift sample are on average larger than that of the non-Swift sample. 

Here our data include GRBs between 28/02/1997-03/05/2008 from Greiner's survey
(\url{http://www.mpe.mpg.de/~jcg}).  The redshift values are taken from this
source, and while $\log T_{90}$ values are also available we do not use the
short-long or the short-intermediate-long grouping \citep{bal98,hor03, varga05,
2008AIPC.1000...56R, hor08,2008AttilaIsmeretlen}. This is beyond the scope of this papers as it
would complicate the statistics due to the different satellites' trigger
criteria. However.  most bursts in our sample come from the long and possibly
the intermediate duration group \citep{hor98,bal99,hor02}.

Note that only satellites triggering GRBs were included here, i.e. we have
ommitted some interesting results regarding the RHESSI satellite
\citep{2006AIPC..836..129R,2006NCimB.121.1493R}.  

Since the Ulysses, ASM and XTR trio observed a total of 8 GRBs, therefore we
aggregated them into one group (labeled Ulysses).  The detailed statistics are
the following: \\ \ \\

$\begin{array}{l|r|r|l|l}
\hline 
\mbox{Spacecraft} & \mbox{GRB} & \mbox{GRB with $z$} & z_{\mbox{min}}& z_{\mbox{max}}
\\ \hline 
\parbox{8.3cm}{HETE}& 79 & 20 & 0.1606 & 3.372 \cr
\parbox{8.3cm}{SAX} & 57 & 19 & 0.0085 & 3.9 \cr
\parbox{8.3cm}{Swift}& 315 & 103 & 0.0331 & 6.29 \cr
\parbox{8.3cm}{\begin{flushleft} Ulysses group (Ulysses + ASM + XTR) \end{flushleft} } & 64 & 8 & 0.706 & 4.5 \cr
\hline 
\end{array} $

\section{Biases}

To compare the $z$ distributions the Swift and non-Swift samples were compared
using non-parametric rank based tests: the Kolmogorov-Smirnov test and the
median test.  These rank based tests have the clear advantage of being
unaffected by any monotonous transformation in the $z$ values.

The Kolmogorov-Smirnov test compares the maximum difference in the cumulative
distributions of the redshifts in the two samples.  
The median test compares the medians of the Swift and non-Swift samples as
follows: be chosen $N_{\mbox{Swift}}$ objects randomly from the sample of the
non-Swift events ($N_{\mbox{Swift}}$ denotes the number of GRBs in the Swift
sample), and calculate the median. Repeat this e.g. 100000 times, and these
Monte-Carlo simulations give the median distribution for $N_{\mbox{Swift}}$
random GRBs selected from the non-Swift group. Comparing this distribution with
the real Swift median $z$ gives us the significance level for the null
hypothesis that the two medians are equal.

The significance  of the Kolmogorov-Smirnov and the median test changes as new
data arrive continuously from the spacecrafts.  The significances show gradual
fall till 10/2006, however after the probabilities rise - while the length of
the datasets grows!  This kind time dependence indicates some fundamental
change in the global observational strategy.

There are definite selections effects from satellite lifespan and sky coverage
E.g. the Swift's X-ray afterglow observations revisit the earlier GRB
directions and create hot spots in the GRB sky distributions.  The optical
follow-up observations' sky coverage is strongly biased biased too, and it
changes from spacecraft to spacecraft. It is due to the different technical
limitations, telescope aviability and the scientific community interest.

\begin{figure}
\caption{ SAX's 
redshift-galactic latitude  and 
and  Swift's 
redshift-declination distribution. 
The galactic disk is clearly visible as a void between $-10 < b < 10$. 
There are some signs of the north/south asymmetry, too.}
\includegraphics[width=0.5\textwidth]{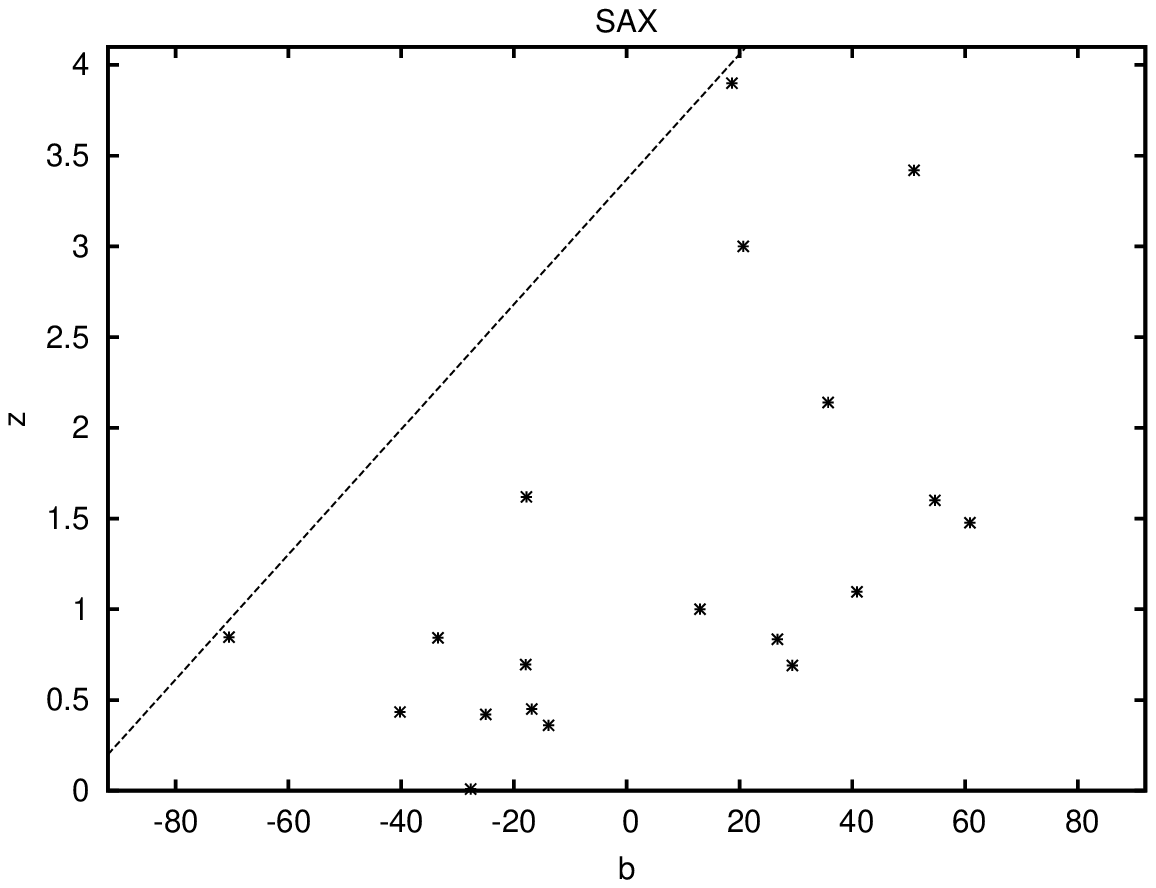}
\includegraphics[width=0.5\textwidth]{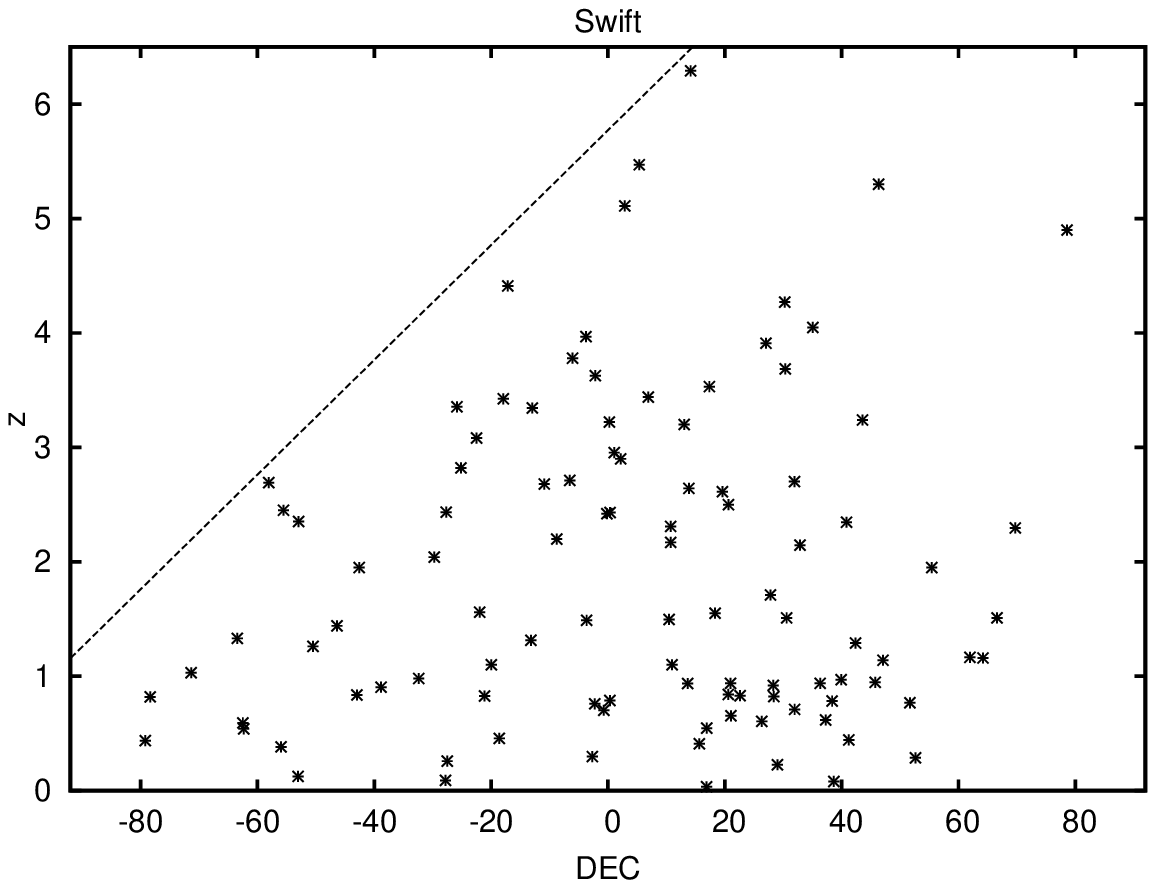}
\label{sax}
\end{figure}

On Fig. \ref{sax}. we show the non-isotropic redshift distribution
of the SAX's and Swift's GRBs: both in the galactic and in the equatorial
system there are strong selection effects. The galactic disk is clearly visible
as a void around $-10 < b < 10 $, and the clear cutoff in the redshift at low
declinations shows that the majority of the optical observations were made on
the northern hemisphere.

\section{Reconstruction} 

The observational biases demonstrated in the previous section can be accounted
for - the reconstructions are similar to the magnitude limited quasar sample.
We used a reconstruction technique based on the Lynden-Bell's C- method
\citep{1971MNRAS.155...95L}, \cite{1987MNRAS.226..273C} to generate weights
from the untruncated part of the data and reconstruct the original
(untruncated) density function.  

\begin{figure}
\caption{The raw $n(<z)$ cumulative distribution of the different spacecrafts' GRB observations and the reconstructed $n(<z)$.} 
\includegraphics[height=0.5\textwidth, angle=270]{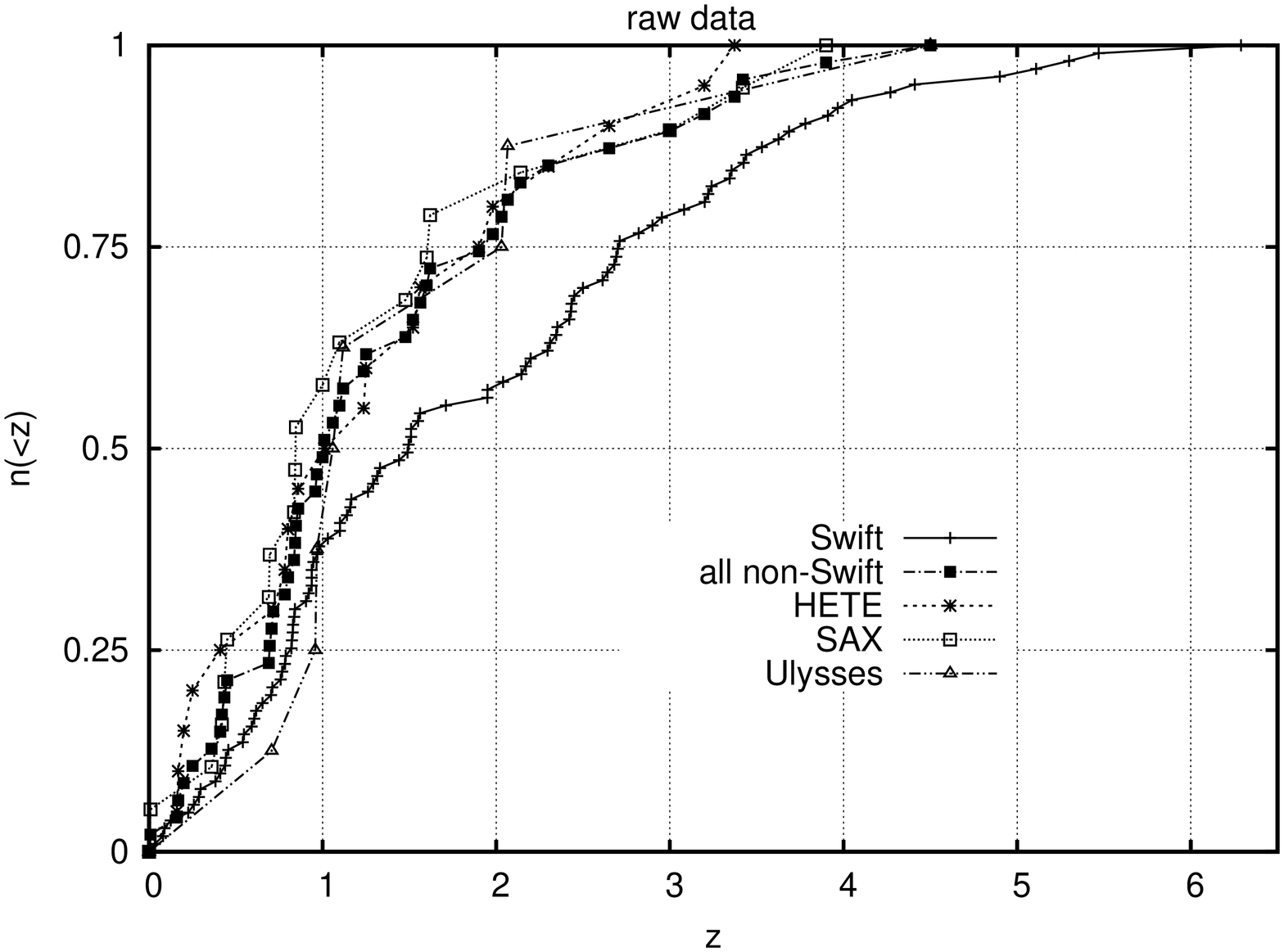}
\includegraphics[height=0.5\textwidth, angle=270]{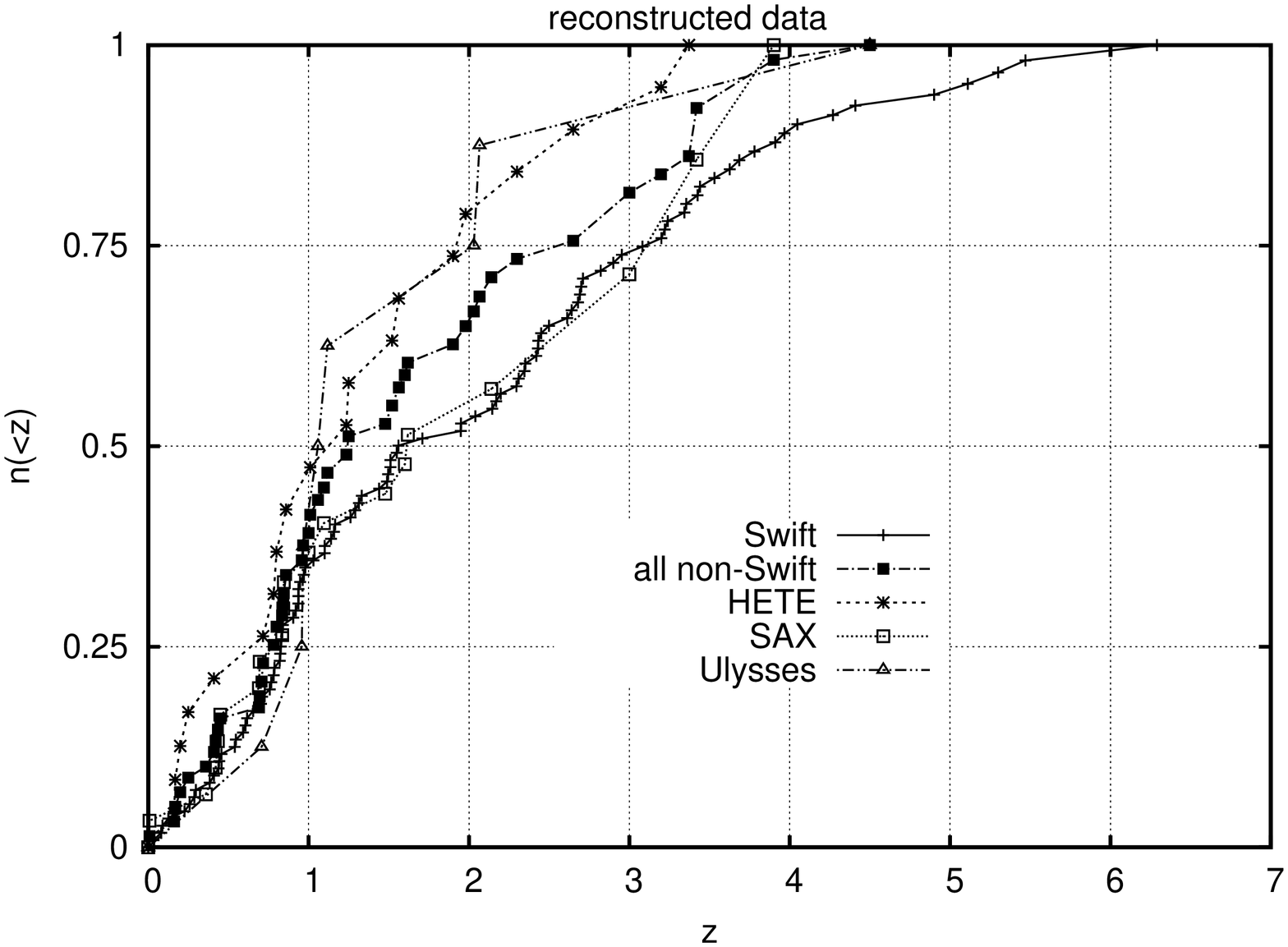} 
\label{reconst}
\end{figure}

Let sort our bursts in ascending order by $z$, and let $\Omega_{i}$ be the
solid angle where all bursts with $z<z_i$ can be detected. Here $\Omega_{i+1}
\subseteq \Omega_i$, which simplifies the analysis.  

We construct the real $n(<z)$ cumulative density function in the following way:
let $N_i= \sum_{j\in\Omega_i} 1 $, i.e. there are $N_i$ burst within the
$\Omega_{i}$ region. Here $n(<z_{i})$ is untruncated, hence $n(<z_{i+1}) =
n(<z_{i}) (N_i + 1 ) / N_i $.  Starting the sequence with $n(<z_1)=1$ we can
reconstruct the cumulative density function.

For the $\Omega_{z}$ sequences we considered both the $b$ and declination cuts
for each spacecrafts, determined from the real observational data.  
Fig.\ref{reconst}. shows the reconstructed $n(<z)$ cumulative distribution of the different spacecrafts' GRB observations.
Here the KS test gives $p=95.7\%$ for the HETE and Swift distribution.


\begin{theacknowledgments}
This research is supported from Hungarian OTKA grant T048870, and partially from ELTE eScience RET and a Pol\'anyi grant (KFKT-2006-01-00012) (P.V.).
\end{theacknowledgments}


\bibliographystyle{aipproc}
\bibliography{zbagoly_proc}

\end{document}